# On Non-Monochromatic Plane-Wave Solutions of Maxwell's Equations

**Costas J. Papachristou**

Department of Physical Sciences, Hellenic Naval Academy, Piraeus, Greece

**Abstract.** In electrodynamics courses and textbooks the properties of plane electromagnetic waves in both conducting and non-conducting media are typically studied from the point of view of the prototype case of a monochromatic plane wave. In this article the more general, non-monochromatic case is examined, which better exploits the independence of the Maxwell equations and offers new insights in the study of electromagnetic waves.

## Introduction

Plane electromagnetic (e/m) waves constitute a significant type of solution of the time-dependent Maxwell equations. A standard textbook approach, both at the intermediate level (e.g., Griffiths, 2013; Wangsness, 1986; Papachristou, 2020) and the advanced level (e.g., Jackson, 1999; Greiner, 1998), is to study the properties of the prototype monochromatic plane wave in a conducting or a non-conducting medium. This article examines the more general and much more interesting case of non-monochromatic plane e/m waves.

Relatively recently the Maxwell system of equations was shown to be a sort of *Bäcklund transformation* (BT for short; see Appendix) relating the e/m wave equations for the electric and the magnetic field (Papachristou, 2015 and 2019; Papachristou and Magoulas, 2016). Considered as a BT, the Maxwell system is a set of first-order partial differential equations (PDEs) the self-consistency of which set requires that the electric and the magnetic field separately satisfy a certain higher-order PDE, namely, a wave equation whose form depends on the medium of propagation of the e/m wave. Technically speaking, the wave equations for the electric and the magnetic field are *integrability conditions* (or *consistency conditions*) in order for the BT to be integrable for either field when the other field is known. We say that the solutions of the wave equations for the electric and the magnetic field, which jointly satisfy the Maxwell system, are *conjugate* through the BT.

A method for finding pairs $(\vec{E},\vec{B})$ of conjugate solutions is the following: Since the wave equation is linear, it is not hard to find general, parameter-dependent solutions for the *E*-field and the *B*-field. In order that these solutions together satisfy the BT, the parameters of the two fields must be appropriately related. To establish this relation we substitute the parametric solutions into the BT (that is, into the Maxwell system) and find the necessary conditions in order that the *E*-field parameters and the *B*-field parameters match. The parametric pair $(\vec{E},\vec{B})$, then, satisfies the Maxwell system and represents an e/m wave.

The simplest application of this method concerns the well-known case of a monochromatic plane wave of angular frequency $\omega$ (see, e.g., Papachristou and Magoulas, 2016). The frequency $\omega$ together with the wave vector and the amplitudes of the $E$ and $B$-fields constitute the "parameters" of the problem, which must be properly related in order that the pair $(\vec{E},\vec{B})$ satisfy the Maxwell system.

The choice of simple monochromatic solutions, however, does not fully exploit the independence of the Maxwell equations (Papachristou and Magoulas, 2022). In this article the more general, non-monochromatic case is studied that makes minimal initial assumptions regarding the specific functional forms of the plane waves representing the electric and the magnetic field. The only assumption one does need to make from the outset is that both fields (electric and magnetic) are expressible in Fourier-integral form as linear superpositions of monochromatic waves of various frequencies. In particular, it is not even necessary to *a priori* require that the plane waves representing the two fields travel in the same direction.

In the next section we review the case of a monochromatic plane e/m wave in empty space. A more general, non-monochromatic treatment of the plane-wave-propagation problem in empty space is then described and this approach is extended to plane-wave solutions for a conducting medium. An interesting difference from the monochromatic case is noted in this regard.

## Monochromatic wave in empty space

In empty space where no charges or currents (free or bound) exist, the Maxwell equations are a system of homogeneous linear first-order PDEs:

$$\begin{aligned} &(a)\quad \vec{\nabla}\cdot\vec{E}=0 \qquad &&(c)\quad \vec{\nabla}\times\vec{E}=-\frac{\partial\vec{B}}{\partial t} \\ &(b)\quad \vec{\nabla}\cdot\vec{B}=0 \qquad &&(d)\quad \vec{\nabla}\times\vec{B}=\varepsilon_0\mu_0\frac{\partial\vec{E}}{\partial t} \end{aligned} \tag{1}$$

where $\vec{E}$ and $\vec{B}$ are the electric and the magnetic field, respectively. This system may be viewed as a Bäcklund transformation (BT; see Appendix) relating solutions of the second-order wave equations for the electric and the magnetic field. This means that the self-consistency of the system (1) requires that each field $\vec{E}$ and $\vec{B}$ satisfy a corresponding wave equation. The wave equations for $\vec{E}$ and $\vec{B}$ are thus consistency conditions for the BT (1). This can be proven by applying the vector identities

$$\vec{\nabla}\times(\vec{\nabla}\times\vec{E})=\vec{\nabla}(\vec{\nabla}\cdot\vec{E})-\nabla^2\vec{E}$$

$$\vec{\nabla}\times(\vec{\nabla}\times\vec{B})=\vec{\nabla}(\vec{\nabla}\cdot\vec{B})-\nabla^2\vec{B}$$

to the system (1), by which process we obtain separate wave equations for $\vec{E}$ and $\vec{B}$:

$$\nabla^2\vec{E}-\frac{1}{c^2}\frac{\partial^2\vec{E}}{\partial t^2}=0 \tag{2}$$

$$\nabla^2\vec{B}-\frac{1}{c^2}\frac{\partial^2\vec{B}}{\partial t^2}=0 \tag{3}$$

where $c$ is the speed of propagation of an e/m wave in vacuum, equal to

$$c=\frac{1}{\sqrt{\varepsilon_0\mu_0}} \tag{4}$$

As mentioned earlier, the fields $\vec{E}$ and $\vec{B}$ related by (1) are *conjugate* through the BT (1). To find pairs of conjugate fields we seek parametric solutions of the PDEs (2) and (3) (see Appendix). To this end we try monochromatic plane-wave solutions of angular frequency $\omega$, propagating in the direction of the wave vector $\vec{k}$ :

$$\begin{aligned} &\vec{E}(\vec{r},t)=\vec{E}_0\exp\{\,i\,(\vec{k}\cdot\vec{r}-\omega t)\} \qquad (a)\\ &\vec{B}(\vec{r},t)=\vec{B}_0\exp\{\,i\,(\vec{k}\cdot\vec{r}-\omega t)\} \qquad (b)\end{aligned} \tag{5}$$

where $\vec{E}_0$ and $\vec{B}_0$ are constant complex amplitudes and where

$$\frac{\omega}{k}=c \qquad (k=|\vec{k}\,|) \tag{6}$$

The $\omega$, $\vec{k}$ , $\vec{E}_0$ and $\vec{B}_0$ are the "parameters" on which the test solutions (5) depend.

The general solutions (5) are not *a priori* conjugate through the BT (1) and hence do not represent an e/m field. To find the extra constraints that must be satisfied by the parameters we need to substitute Eqs. (5) into the Maxwell system (1). By taking into account that $\vec{\nabla}e^{\,i\vec{k}\cdot\vec{r}}=i\vec{k}\,e^{\,i\vec{k}\cdot\vec{r}}$, the *div* equations (1*a*) and (1*b*) yield

$$\vec{k}\cdot\vec{E}_0=0 \quad (a) \qquad \vec{k}\cdot\vec{B}_0=0 \quad (b) \tag{7}$$

while the *rot* equations (1*c*) and (1*d*) give

$$\vec{k}\times\vec{E}_0=\omega\,\vec{B}_0 \quad (a) \qquad \vec{k}\times\vec{B}_0=-\frac{\omega}{c^2}\,\vec{E}_0 \quad (b) \tag{8}$$

Now, we notice that the four equations (7)–(8) do not form an independent set since (7*b*) and (8*b*) can be reproduced by using (7*a*) and (8*a*). Indeed, taking the dot product of (8*a*) with $\vec{k}$ we get (7*b*), while taking the cross product of (8*a*) with $\vec{k}$ , and using (7*a*) and (6), we find (8*b*).

So, from 4 independent Maxwell equations we obtained only 2 independent pieces of information. This happened because we "fed" our trial solutions (5) with more information than necessary, in anticipation of results that follow *a posteriori* from Maxwell's equations. Thus, we assumed from the outset that the two waves (electric and magnetic) have similar simple functional forms and propagate in the same direction. By relaxing these initial assumptions our analysis acquires a richer and more interesting structure.

## Non-monochromatic wave in empty space

Let us assume, more generally, that the fields $\vec{E}$ and $\vec{B}$ represent plane waves propagating in empty space in the directions of the unit vectors $\hat{\tau}$ and $\hat{\kappa}$, respectively:

$$\vec{E}(\vec{r},t)=\vec{F}(\hat{\tau}\cdot\vec{r}-ct)\ ,\quad \vec{B}(\vec{r},t)=\vec{G}(\hat{\kappa}\cdot\vec{r}-ct) \tag{9}$$

Furthermore, we assume that the functions $\vec{F}$ and $\vec{G}$ can be expressed as linear combinations of monochromatic plane waves of the form (5), for continuously varying values of $k$ and $\omega$, where $\omega=ck$, according to (6). Then $\vec{E}$ and $\vec{B}$ can be written in Fourier-integral form, as follows:

$$\begin{aligned}\vec{E}&=\int \vec{E}_0(k)\,e^{ik(\hat{\tau}\cdot\vec{r}-ct)}dk\\ \vec{B}&=\int \vec{B}_0(k)\,e^{ik(\hat{\kappa}\cdot\vec{r}-ct)}dk\end{aligned} \tag{10}$$

In general, the integration variable $k$ is assumed to run from 0 to $+\infty$. For notational economy, the limits of integration with respect to $k$ will not be displayed explicitly.

By setting

$$u=\hat{\tau}\cdot\vec{r}-ct\ ,\quad v=\hat{\kappa}\cdot\vec{r}-ct \tag{11}$$

we write

$$\begin{aligned}\vec{E}(u)&=\int \vec{E}_0(k)\,e^{iku}dk\\ \vec{B}(v)&=\int \vec{B}_0(k)\,e^{ikv}dk\end{aligned} \tag{12}$$

We note that

$$\vec{\nabla}e^{iku}=ik\hat{\tau}e^{iku}\ ,\quad \vec{\nabla}e^{ikv}=ik\hat{\kappa}e^{ikv} \tag{13}$$

By using (12) and (13) we find that

$$\vec{\nabla}\cdot\vec{E}=\int ik\hat{\tau}\cdot\vec{E}_0(k)\,e^{iku}dk\ ,\quad \vec{\nabla}\cdot\vec{B}=\int ik\hat{\kappa}\cdot\vec{B}_0(k)\,e^{ikv}dk\ ,$$

$$\vec{\nabla}\times\vec{E}=\int ik\hat{\tau}\times\vec{E}_0(k)\,e^{iku}dk\ ,\quad \vec{\nabla}\times\vec{B}=\int ik\hat{\kappa}\times\vec{B}_0(k)\,e^{ikv}dk\ .$$

Moreover, we have that

$$\frac{\partial\vec{E}}{\partial t}=-\int i\omega\vec{E}_0(k)\,e^{iku}dk\ ,\quad \frac{\partial\vec{B}}{\partial t}=-\int i\omega\vec{B}_0(k)\,e^{ikv}dk$$

where, as always, $\omega=ck$.

The two Gauss' laws (1*a*) and (1*b*) yield

$$\int k\hat{\tau}\cdot\vec{E}_0(k)\,e^{iku}dk=0 \quad \text{and} \quad \int k\hat{\kappa}\cdot\vec{B}_0(k)\,e^{ikv}dk=0\,,$$

respectively. In order that these relations be valid identically for all *u* and all *v*, respectively, we must have

$$\hat{\tau}\cdot\vec{E}_0(k)=0 \quad \text{and} \quad \hat{\kappa}\cdot\vec{B}_0(k)=0\,, \ \text{ for all } k \tag{14}$$

From Faraday's law (1*c*) and the Ampère-Maxwell law (1*d*) we obtain two more integral equations:

$$\int k\hat{\tau}\times\vec{E}_0(k)\,e^{iku}dk=\int \omega\vec{B}_0(k)\,e^{ikv}dk \tag{15}$$

$$\int k\hat{\kappa}\times\vec{B}_0(k)\,e^{ikv}dk=-\int \frac{\omega}{c^2}\vec{E}_0(k)\,e^{iku}dk \tag{16}$$

where we have taken into account Eq. (4).

Taking the cross product of (15) with $\hat{\kappa}$ and using (16), we find the integral relation

$$\int k\,[(\hat{\kappa}\cdot\vec{E}_0)\,\hat{\tau}-(\hat{\kappa}\cdot\hat{\tau})\vec{E}_0]\,e^{iku}dk=-\int k\vec{E}_0\,e^{iku}dk\,.$$

This is true for all *u* if

$$(\hat{\kappa}\cdot\vec{E}_0)\,\hat{\tau}-(\hat{\kappa}\cdot\hat{\tau})\vec{E}_0=-\vec{E}_0 \ \Rightarrow\ (\hat{\kappa}\cdot\hat{\tau}-1)\vec{E}_0=(\hat{\kappa}\cdot\vec{E}_0)\,\hat{\tau}\,.$$

Given that, by (14), $\vec{E}_0$ and $\hat{\tau}$ are mutually perpendicular, the above relation can only be valid if $\hat{\kappa}\cdot\hat{\tau}=1$ and $\hat{\kappa}\cdot\vec{E}_0=0$. This, in turn, can only be satisfied if $\hat{\kappa}=\hat{\tau}$. The same conclusion is reached by taking the cross product of (16) with $\hat{\tau}$ and by using (15) as well as the fact that $\vec{B}_0$ is normal to $\hat{\kappa}$. From (11) we then have that

$$u=v=\hat{\tau}\cdot\vec{r}-ct$$

so that relations (12) become

$$\begin{aligned}\vec{E}(\vec{r},t)&=\int \vec{E}_0(k)\,e^{iku}dk=\int \vec{E}_0(k)\,e^{ik(\hat{\tau}\cdot\vec{r}-ct)}dk\\ \vec{B}(\vec{r},t)&=\int \vec{B}_0(k)\,e^{iku}dk=\int \vec{B}_0(k)\,e^{ik(\hat{\tau}\cdot\vec{r}-ct)}dk\end{aligned} \tag{17}$$

Equations (14) are now rewritten as

$$\hat{\tau}\cdot\vec{E}_0(k)=0 \quad \text{and} \quad \hat{\tau}\cdot\vec{B}_0(k)=0\,, \ \text{ for all } k \tag{18}$$

Furthermore, in order that (15) and (16) (with *u* and $\hat{\tau}$ in place of *v* and $\hat{\kappa}$, respectively) be identically valid for all *u*, we must have

$$k\,\hat{\tau}\times\vec{E}_0(k)=\omega\vec{B}_0(k)\;\Leftrightarrow\;\hat{\tau}\times\vec{E}_0(k)=c\vec{B}_0(k) \tag{19}$$

and

$$k\,\hat{\tau}\times\vec{B}_0(k)=-\frac{\omega}{c^2}\vec{E}_0(k)\;\Leftrightarrow\;\hat{\tau}\times\vec{B}_0(k)=-\frac{1}{c}\vec{E}_0(k) \tag{20}$$

for all $k$, where $k=\omega/c$. Notice, however, that (19) and (20) are not independent equations, since (20) is essentially the cross product of (19) with $\hat{\tau}$.

The general plane-wave solutions to the Maxwell system (1) are thus given by relations (17) with the additional constraints (18) and (19) on the parameters. Let us summarize our main findings:

1. The fields $\vec{E}$ and $\vec{B}$ are plane waves traveling in the same direction, defined by the unit vector $\hat{\tau}$; these fields satisfy the Maxwell equations in empty space.

2. The e/m wave $(\vec{E},\vec{B})$ is a *transverse* wave. Indeed, from equations (17) and the orthogonality relations (18) it follows that

$$\hat{\tau}\cdot\vec{E}=0 \quad\text{and}\quad \hat{\tau}\cdot\vec{B}=0 \tag{21}$$

3. The fields $\vec{E}$ and $\vec{B}$ are mutually perpendicular. Moreover, the vectors $(\vec{E},\vec{B},\hat{\tau})$ define a right-handed rectangular system. Indeed, by cross-multiplying (17) with $\hat{\tau}$ and by using (19) and (20), we find:

$$\hat{\tau}\times\vec{E}=c\vec{B}\;,\quad \hat{\tau}\times\vec{B}=-\frac{1}{c}\vec{E} \tag{22}$$

4. Taking *real values* of (21) and (22), we have:

$$\hat{\tau}\cdot\mathrm{Re}\vec{E}=0\;,\quad \hat{\tau}\cdot\mathrm{Re}\vec{B}=0 \quad\text{and}\quad \hat{\tau}\times\mathrm{Re}\vec{E}=c\,\mathrm{Re}\vec{B} \tag{23}$$

The magnitude of the last vector equation in (23) gives a relation between the instantaneous values of the electric and the magnetic field:

$$|\,\mathrm{Re}\vec{E}\,|=c\,|\,\mathrm{Re}\vec{B}\,| \tag{24}$$

The above results for empty space can be extended in a straightforward way to the case of a *linear, non-conducting, non-dispersive* medium upon replacement of $\varepsilon_0$ and $\mu_0$ with $\varepsilon$ and $\mu$, respectively (see, e.g., Papachristou, 2020). The (frequency-independent) speed of propagation of the plane e/m wave in this case is $\upsilon=1/(\varepsilon\mu)^{1/2}$.

## Plane wave in a conducting medium

The Maxwell equations for a conducting medium of conductivity $\sigma$ may be written as follows (Griffiths, 2013; Papachristou, 2020):

$$
\begin{aligned}
&(a) \quad \vec{\nabla}\cdot\vec{E}=0 \qquad && (c) \quad \vec{\nabla}\times\vec{E}=-\frac{\partial\vec{B}}{\partial t} \\
&(b) \quad \vec{\nabla}\cdot\vec{B}=0 \qquad && (d) \quad \vec{\nabla}\times\vec{B}=\mu\sigma\vec{E}+\varepsilon\mu\frac{\partial\vec{E}}{\partial t}
\end{aligned}
\tag{25}
$$

By using the vector identities

$$\vec{\nabla}\times(\vec{\nabla}\times\vec{E})=\vec{\nabla}(\vec{\nabla}\cdot\vec{E})-\nabla^2\vec{E}$$

$$\vec{\nabla}\times(\vec{\nabla}\times\vec{B})=\vec{\nabla}(\vec{\nabla}\cdot\vec{B})-\nabla^2\vec{B}$$

the relations (25) lead to the *modified wave equations*

$$\nabla^2\vec{E}-\varepsilon\mu\frac{\partial^2\vec{E}}{\partial t^2}-\mu\sigma\frac{\partial\vec{E}}{\partial t}=0 \tag{26}$$

$$\nabla^2\vec{B}-\varepsilon\mu\frac{\partial^2\vec{B}}{\partial t^2}-\mu\sigma\frac{\partial\vec{B}}{\partial t}=0 \tag{27}$$

Guided by our monochromatic-wave approach to the problem (see, e.g., Papachristou and Magoulas, 2016) we now try a more general, integral form of solution of the above wave equations:

$$
\begin{aligned}
\vec{E}(\vec{r},t)&=\int\vec{E}_0(k)\,e^{-s\hat{\tau}\cdot\vec{r}}e^{i(k\,\hat{\tau}\cdot\vec{r}-\omega t)}dk=\int\vec{E}_0(k)\exp\{(ik-s)\hat{\tau}\cdot\vec{r}-i\omega t\}\,dk \\
\vec{B}(\vec{r},t)&=\int\vec{B}_0(k)\,e^{-s\hat{\tau}\cdot\vec{r}}e^{i(k\,\hat{\tau}\cdot\vec{r}-\omega t)}dk=\int\vec{B}_0(k)\exp\{(ik-s)\hat{\tau}\cdot\vec{r}-i\omega t\}\,dk
\end{aligned}
\tag{28}
$$

where $s$ is a real parameter related to the conductivity of the medium. As in the vacuum case, the unit vector $\hat{\tau}$ indicates the direction of propagation of the wave. Notice that we have assumed from the outset that both waves – electric and magnetic – propagate in the same direction, in view of the fact that our results must agree with those for a non-conducting medium (in particular, for the vacuum) upon setting $s$=0.

It is convenient to set

$$\exp\{(ik-s)\,\hat{\tau}\cdot\vec{r}-i\omega t\}\equiv A(\vec{r},t) \tag{29}$$

Then, Eq. (28) takes on the form

$$
\begin{aligned}
\vec{E}(\vec{r},t)&=\int\vec{E}_0(k)\,A(\vec{r},t)\,dk \\
\vec{B}(\vec{r},t)&=\int\vec{B}_0(k)\,A(\vec{r},t)\,dk
\end{aligned}
\tag{30}
$$

The following relations can be proven:

$$\vec{\nabla} A(\vec{r},t)=(ik-s)\,\hat{\tau} A(\vec{r},t) \tag{31}$$

$$\nabla^2 A(\vec{r},t)=(s^2-k^2-2isk)A(\vec{r},t) \tag{32}$$

Moreover,

$$\frac{\partial}{\partial t}A(\vec{r},t)=-i\omega A(\vec{r},t) \quad \text{and} \quad \frac{\partial^2}{\partial t^2}A(\vec{r},t)=-\omega^2 A(\vec{r},t)\,.$$

From (26) we get

$$\int [(s^2-k^2+\varepsilon\mu\omega^2)+i(\mu\sigma\omega-2sk)]\,\vec{E}_0(k)A(\vec{r},t)\,dk=0$$

[a similar integral relation is found from (27)]. This will be identically satisfied for all $\vec{r}$ and $t$ if

$$s^2-k^2+\varepsilon\mu\omega^2=0 \quad \text{and} \quad \mu\sigma\omega-2sk=0 \tag{33}$$

By using relations (33), $\omega$ and $s$ can be expressed as functions of $k$, as required in order that the integral relations (28) make sense. Notice, in particular, that, by the second relation (33), $s=0$ if $\sigma=0$ (non-conducting medium). Then, by the first relation, $\omega/k=1/(\varepsilon\mu)^{1/2}$, which is the familiar expression for the speed of propagation of an e/m wave in a non-conducting medium.

From the two Gauss' laws (25*a*) and (25*b*) we get the corresponding integral relations

$$\int (ik-s)\,\hat{\tau}\cdot\vec{E}_0(k)A(\vec{r},t)\,dk=0\,,$$
$$\int (ik-s)\,\hat{\tau}\cdot\vec{B}_0(k)A(\vec{r},t)\,dk=0\,.$$

These will be identically satisfied for all $\vec{r}$ and $t$ if

$$\hat{\tau}\cdot\vec{E}_0(k)=0 \quad \text{and} \quad \hat{\tau}\cdot\vec{B}_0(k)=0, \ \text{for all } k \tag{34}$$

From (25*c*) and (25*d*) we find

$$\int (ik-s)\,\hat{\tau}\times\vec{E}_0(k)A(\vec{r},t)\,dk=\int i\omega\vec{B}_0(k)A(\vec{r},t)\,dk$$

and

$$\int (ik-s)\,\hat{\tau}\times\vec{B}_0(k)A(\vec{r},t)\,dk=\int (\mu\sigma-i\varepsilon\mu\omega)\vec{E}_0(k)A(\vec{r},t)\,dk\,,$$

respectively. To satisfy these for all $\vec{r}$ and $t$, we require that

$$(k+is)\,\hat{\tau}\times\vec{E}_0(k)=\omega\vec{B}_0(k) \tag{35}$$

and

$$(k+is)\,\hat{\tau}\times\vec{B}_0(k)=-(\varepsilon\mu\omega+i\mu\sigma)\vec{E}_0(k) \qquad (36)$$

Note, however, that (36) is not an independent equation since it can be reproduced by cross-multiplying (35) with $\hat{\tau}$ and by taking into account Eqs. (33) and (34).

We note the following:

1. From (30) and (34) we have that

$$\hat{\tau}\cdot\vec{E}=0 \quad \text{and} \quad \hat{\tau}\cdot\vec{B}=0 \qquad (37)$$

or, in real form, $\hat{\tau}\cdot\mathrm{Re}\,\vec{E}=0$ and $\hat{\tau}\cdot\mathrm{Re}\,\vec{B}=0$. This means that both $\mathrm{Re}\,\vec{E}$ and $\mathrm{Re}\,\vec{B}$ are normal to the direction of propagation of the wave.

2. From (30) and (35) we get

$$\hat{\tau}\times\vec{E}=\int\frac{\omega}{k+is}\vec{B}_0(k)A(\vec{r},t)\,dk \qquad (38)$$

The integral on the right-hand side of (38) is, generally, not a vector parallel to $\vec{B}$. Now, in the limit of negligible conductivity ($\sigma$=0) the relations (33) give $s$=0 and $\omega/k=1/(\varepsilon\mu)^{1/2}$. The quotient $\omega/k$ represents the speed of propagation $\upsilon$ in the non-conducting medium, for the frequency $\omega$. If the medium is *non-dispersive*, the speed $\upsilon=\omega/k$ is constant, independent of frequency. Then Eq. (38) (with $s$=0) becomes

$$\hat{\tau}\times\vec{E}=\upsilon\int\vec{B}_0(k)A(\vec{r},t)\,dk=\upsilon\vec{B}$$

and, in real form, it reads $\hat{\tau}\times\mathrm{Re}\vec{E}=\upsilon\,\mathrm{Re}\vec{B}$. Geometrically, this means that the vectors $(\mathrm{Re}\,\vec{E},\mathrm{Re}\,\vec{B},\hat{\tau})$ define a right-handed rectangular system.

3. The vectors $\vec{E}$ and $\vec{B}$ are mutually perpendicular in a *monochromatic* e/m wave of definite frequency $\omega$, traveling in a conducting medium (see, e.g., Papachristou and Magoulas, 2016). Such a wave is represented in real form by the equations

$$\vec{E}(\vec{r},t)=\vec{E}_0\,e^{-s\hat{\tau}\cdot\vec{r}}\cos(k\hat{\tau}\cdot\vec{r}-\omega t+\alpha)\;,$$

$$\vec{B}(\vec{r},t)=\frac{\sqrt{k^2+s^2}}{\omega}\,(\hat{\tau}\times\vec{E}_0)\,e^{-s\hat{\tau}\cdot\vec{r}}\cos(k\hat{\tau}\cdot\vec{r}-\omega t+\beta)$$

where $\vec{E}_0$ is a real vector and where $\tan(\beta-\alpha)=s/k$. This perpendicularity between $\vec{E}$ and $\vec{B}$ ceases to exist, however, in a non-monochromatic wave of the form (28).

## Summary

The Maxwell equations of electrodynamics may be viewed as a Bäcklund transformation (BT) relating solutions of the wave equations for the electric and the magnetic field (Papachristou, 2015; Papachristou and Magoulas, 2016). Wave solutions for the *E* and *B*-fields that jointly satisfy the BT are said to be *conjugate* through the BT.

Pairs of conjugate solutions can be found by assuming general, parameter-dependent solutions of the two wave equations (electric and magnetic) and by seeking the conditions these parameters must obey in order for the Maxwell BT to be satisfied. The usual choice of test solutions is the monochromatic plane e/m wave, the properties of which are, of course, well known.

In this article a more general choice of wave solutions was made that better exploits the independence of the Maxwell equations. It concerns the non-monochromatic plane wave composed of an infinite number of monochromatic waves of various frequencies. The non-monochromatic case exhibits some interesting properties. For example, one does not have to *a priori* assume that the electric and magnetic waves propagate in the same direction: this is dictated by the Maxwell system itself. Moreover, it is only in the monochromatic case that the *E* and *B*-fields of a plane e/m wave in a conducting medium are mutually perpendicular, whereas this is not the case for a non-monochromatic wave in such a medium (in fact, in any *dispersive* medium).

## Appendix. Bäcklund transformations

Consider two PDEs $P[u]=0$ and $Q[v]=0$ for the unknown functions $u$ and $v$, respectively. The expressions $P[u]$ and $Q[v]$ may contain the corresponding variables $u$ and $v$, as well as partial derivatives of $u$ and $v$ with respect to the independent variables. For simplicity, we assume that $u$ and $v$ are functions of only two variables $x$, $y$. Partial derivatives with respect to these variables will be denoted by using subscripts: $u_x$ , $u_y$ , $u_{xx}$ , $u_{yy}$ , $u_{xy}$ , etc.

Independently, for the moment, also consider a pair of coupled PDEs for $u$ and $v$:

$$B_1[u,v]=0 \quad (a) \qquad B_2[u,v]=0 \quad (b) \tag{39}$$

where the expressions $B_i\,[u,v]$ ($i$=1,2) may contain $u$, $v$ as well as partial derivatives of $u$ and $v$ with respect to $x$ and $y$. We note that $u$ appears in both equations ($a$) and ($b$). The question then is: if we find an expression for $u$ by integrating ($a$) for a given $v$, will it match the corresponding expression for $u$ found by integrating ($b$) for the same $v$? The answer is that, in order that ($a$) and ($b$) be consistent with each other for solution for $u$, the function $v$ must be properly chosen so as to satisfy a certain *consistency condition* (or *integrability condition* or *compatibility condition*).

By a similar reasoning, in order that ($a$) and ($b$) in (39) be mutually consistent for solution for $v$, for some given $u$, the function $u$ must now itself satisfy a corresponding integrability condition.

If it happens that the two consistency conditions for integrability of the system (39) are precisely the PDEs $P[u]=0$ and $Q[v]=0$, we say that this system constitutes a *Bäcklund transformation* (BT) connecting solutions of $P[u]=0$ with solutions of $Q[v]=0$. In the special case where $P\equiv Q$, i.e., if $u$ and $v$ satisfy *the same* PDE, the system (39) is called an *auto-Bäcklund* transformation (auto-BT) for this PDE.

Suppose now that we seek solutions of the PDE $P[u]=0$. Assume that we are able to find a BT connecting solutions $u$ of this equation with solutions $v$ of the PDE $Q[v]=0$ (if $P\equiv Q$ , the auto-BT connects solutions $u$ and $v$ of the same PDE) and let $v=v_0(x,y)$ be some known solution of $Q[v]=0$. The BT is then a system of PDEs for the unknown $u$,

$$B_i[u,v_0]=0\ , \quad i=1,2 \tag{40}$$

The system (40) is integrable for $u$, given that the function $v_0$ satisfies *a priori* the required integrability condition $Q[v]=0$. The solution $u$ then of the system satisfies the PDE $P[u]=0$. Thus a solution $u(x,y)$ of the latter PDE is found without actually solving the equation itself, simply by integrating the BT (40) with respect to $u$. Of course, this method will be useful provided that integrating the system (40) for $u$ is simpler than integrating the PDE $P[u]=0$ itself. If the transformation (40) is an auto-BT for the PDE $P[u]=0$, then, starting with a known solution $v_0(x,y)$ of this equation and integrating the system (40) we find another solution $u(x,y)$ of the same equation.

Let us see an example of the use of a BT to generate solutions of a PDE. The Cauchy-Riemann relations of Complex Analysis,

$$u_x=v_y \quad (a) \qquad u_y=-v_x \quad (b) \tag{41}$$

constitute an auto-BT for the Laplace equation,

$$P[w]\equiv w_{xx}+w_{yy}=0 \tag{42}$$

Let us explain this: Suppose we want to solve the system (41) for $u$, for a given choice of the function $v(x,y)$. To see if the PDEs ($a$) and ($b$) match for solution for $u$ we must compare them in some way. We thus differentiate ($a$) with respect to $y$ and ($b$) with respect to $x$ and equate the mixed derivatives of $u$. That is, we apply the integrability condition $(u_x)_y= (u_y)_x$ . In this way we eliminate the variable $u$ and find the condition that must be obeyed by $v(x,y)$:

$$P[v]\equiv v_{xx}+v_{yy}=0\ .$$

Similarly, by using the integrability condition $(v_x)_y= (v_y)_x$ to eliminate $v$ from the system (41), we find the necessary condition in order that this system be integrable for $v$, for a given function $u(x,y)$:

$$P[u]\equiv u_{xx}+u_{yy}=0\ .$$

In conclusion, the integrability of the system (41) with respect to either variable requires that the other variable satisfy the Laplace equation (42).

Let now $v_0(x,y)$ be a known solution of the Laplace equation (42). Substituting $v=v_0$ in the system (41), we can integrate this system with respect to $u$. It is not hard to show (by eliminating $v_0$ from the system) that the solution $u$ will also satisfy the Laplace equation (42). As an example, by choosing the solution $v_0(x,y)=xy$ , we find a new solution $u(x,y)= (x^2-y^2)/2 +C$ .

As presented above, a BT is an auxiliary device for constructing solutions of a PDE from known solutions of the same or another PDE (this method is particularly useful for finding solutions of nonlinear PDEs). A related problem, where solutions of the differential system representing the BT itself are sought, is also of interest. To be specific, assume that we need to integrate a given system of PDEs connecting two functions $u$ and $v$:

$$B_i\,[u,v]=0\ ,\quad i=1,2 \tag{43}$$

Suppose that the integrability of the system for both functions requires that $u$ and $v$ separately satisfy the respective PDEs

$$P[u]=0\quad (a)\qquad\quad Q[v]=0\quad (b) \tag{44}$$

That is, the system (43) is a BT connecting solutions of the PDEs (44). Assume, now, that the latter PDEs possess known (or, in any case, easy to find) parameter-dependent solutions of the form

$$u=f\,(x,y\,;\alpha,\beta,\ldots)\ ,\quad v=g(x,y\,;\kappa,\lambda,\ldots)$$

where $\alpha$, $\beta$, $\kappa$, $\lambda$, etc., are (real or complex) parameters. If values of these parameters can be determined for which $u$ and $v$ jointly satisfy the system (43), we say that the solutions $u$ and $v$ of the PDEs (44*a*) and (44*b*), respectively, are *conjugate* through the BT (43) (or BT-conjugate, for short). By finding a pair of BT-conjugate solutions one thus automatically obtains a solution of the system (43).

Note that parametric solutions of *both* integrability conditions $P[u]=0$ and $Q[v]=0$ must be known in advance! From the practical point of view the method is thus most applicable in linear problems, since it is much easier to find parameter-dependent solutions of the PDEs (44) in this case.

Let us see an example: Going back to the Cauchy-Riemann relations (41), we try the following parametric solutions of the Laplace equation (42):

$$u\,(x,y)=\alpha\,(x^2-y^2)+\beta x+\gamma y\ ,$$
$$v\,(x,y)=\kappa xy+\lambda x+\mu y\ .$$

Substituting these into the BT (41), we find that $\kappa=2\alpha$, $\mu=\beta$ and $\lambda=-\gamma$. Therefore, the solutions

$$u\,(x,y)=\alpha\,(x^2-y^2)+\beta x+\gamma y\ ,$$
$$v\,(x,y)=2\alpha xy-\gamma x+\beta y$$

of the Laplace equation are BT-conjugate through the Cauchy-Riemann relations.

As a counter-example, let us try a different combination:

$$u\,(x,y)=\alpha xy\ ,\quad v\,(x,y)=\beta xy\ .$$

Inserting these into the system (41) and taking into account the independence of $x$ and $y$, we find that the only possible values of the parameters $\alpha$ and $\beta$ are $\alpha=\beta=0$, so that $u(x,y)= v(x,y)=0$. Thus, no non-trivial BT-conjugate solutions exist in this case.

## References

Greiner W, *Classical Electrodynamics* (Springer, 1998).

Griffiths DJ, *Introduction to Electrodynamics*, 4th Edition (Pearson, 2013).

Jackson JD, *Classical Electrodynamics*, 3rd Edition (Wiley, 1999).

Papachristou CJ, *The Maxwell equations as a Bäcklund transformation*, Advanced Electromagnetics, Vol. 4, No. 1 (2015) 52.[1]

Papachristou CJ, *Aspects of Integrability of Differential Systems and Fields: A Mathematical Primer for Physicists* (Springer, 2019).[2]

Papachristou CJ, *Introduction to Electromagnetic Theory and the Physics of Conducting Solids* (Springer, 2020).[3]

Papachristou CJ, Magoulas AN, *Bäcklund transformations: Some old and new perspectives*, Nausivios Chora, Vol. 6 (2016) C-3.[4]

Papachristou CJ, Magoulas AN, *Independence of Maxwell's equations: A Bäcklund-transformation view*, Nausivios Chora, Vol. 8 (2022) C-3.[5]

Wangsness RK, *Electromagnetic Fields*, 2nd Edition (Wiley, 1986).

[1] https://aemjournal.org/index.php/AEM/article/view/311
[2] https://arxiv.org/abs/1511.01788
[3] https://arxiv.org/abs/1711.09969
[4] https://nausivios.hna.gr/docs/2016C.pdf
[5] https://nausivios.hna.gr/docs/NCH_v8_2022_C1.pdf